\documentstyle[12pt]{article}
\begin{document}
\begin{center}
{\Large \bf Superluminary Universe: A Possible Solution to the Initial
Value Problem in Cosmology}
\vskip 0.3 true in
{\large J. W. Moffat}
\date{}
\vskip 0.3 true in
{\it Department of Physics, University of Toronto,
Toronto, Ontario M5S 1A7, Canada}

\end{center}

\begin{abstract}%
The spontaneous breaking of local Lorentz invariance in the early Universe,
associated with a first order phase transition at a critical time $t_c$,
generates a large increase in the speed of light and a superluminary 
communication of information occurs,
allowing all regions in the Universe to be causally connected. This solves the
horizon problem, leads to a mechanism of monopole suppression in cosmology
and can resolve the flatness problem. After the critical time $t_c$, local 
Lorentz (and diffeomorphism) invariance is 
restored and light travels at its presently measured speed. The kinematical
and dynamical aspects of the generation of quantum fluctuations in the
superluminary Universe are investigated. A scale invariant prediction for
the fluctuation density amplitude is obtained.
\end{abstract}
\vskip 0.5 true in
{\bf Published in: International Journal of Modern Physics D, Vol. 2, No. 3 (1993) 351-365.}
\vskip 0.5 true in
{\bf UTPT-92-14. e-mail: moffat@medb.physics.utoronto.ca}

\section{\bf Introduction}

In order to solve the problem of the initial value of the Universe, popularly
expressed in terms of the standard three problems: the horizon problem,
the flatness problem and the monopole or relic particle suppression
problem, the idea of an inflationary Universe was introduced\cite{Guth,Turner}. By
allowing for an exponential growth of the very early Universe, caused by a
huge vacuum energy generating a de Sitter Universe phase, these initial value 
problems could be solved. In addition, the inflationary mechanism was able
to predict the existence of quantum fluctuations which could be the seeds of
Galaxy formation.
However, the inflationary scenario ran into difficulties already in the
first paper by Guth, since in order to have sufficient inflation, it is 
necessary to keep the ``false" vacuum reasonably stable, leading to a
small bubble nucleation rate and to the lack of bubble coalescence.
The final configuration of the Universe is very inhomogeneous in disagreement
with the observations. This led to modified models introducing special
forms for the Higgs potential $V(\phi)$ with an extremely flat $V(\phi)$
for $\phi \leq \sigma$ ($\sigma =<\phi>_0$)\cite{Turner}. These models used weakly
first order or even second order phase transitions. Instead of many bubbles
colliding and having to coalesce, the models have one huge bubble
containing everything observable in the Universe now. A serious
problem with this kind of model is that the density perturbations,
$\delta\rho/\rho$, are too large by a factor of order $10^6$, unless
the parameter in the potential (e.g. Coleman-Weinberg) is artificially
reduced by a factor $\sim 10^{12}$. 

In Linde's chaotic inflation\cite{Turner}, which does not involve any phase 
transitions,
the potential has the simple form: $V(\phi)=\lambda\phi^4$. Assuming the 
slow motion of $\phi$ from some initial value $\phi_0$ towards the minimum,
one can obtain sufficient inflation. The main difficulties are: 1) To obtain
observationally acceptable values for the density perturbations, it is
necessary to fine-tune $\lambda$ to very small values: $\lambda \approx
10^{-12}-10^{-14}$. There is no obvious physical reason why the coupling
constant $\lambda$ should be so small. 2) It must be assumed that the kinetic
energy of the scalar
field is small compared to its potential energy, which leads to the condition 
that the field be uniform over sizes bigger than the Hubble radius, which
is not in keeping with the original ideas of inflation. 

Other scenarios include
models based on the Brans-Dicke theory of gravity (extended inflation and
hyper-extended inflation) and extensions of this theory\cite{Turner}. These models
also have their problems e.g. by introducing unacceptable distortions of the
microwave background. Up till the present time, there appears to be no
satisfactory model of inflation.

In the following, we shall consider a new scenario which is capable of solving
the horizon, excess relic particle and flatness problems and leads to 
predictions
for small scale inhomogeneities. In previous work\cite{Moffat}, the local Lorentz and
diffeomorphism symmetries of Einstein's gravitational theory are spontaneously
broken by a Higgs mechanism in the early Universe, with the symmetry 
breaking pattern: $SO(3,1)\rightarrow O(3)$, at a critical temperature
$T_c$, below which the symmetry is restored. 

Within this symmetry breaking scheme, the super-Hamiltonian constraint
for the wave function, $H_0\Psi=0$, was relaxed, leading to a time
dependent Schr\"odinger equation. In this framework, quantum mechanics
and gravitation were united in a meaningful observational scheme. The local
Lorentz invariant structure of the gravitational Lagrangian is maintained as
a ``hidden" symmetry. After the spacetime symmetries are restored in the
early Universe for a temperature $T < T_c$, the wave function has an 
oscillatory behavior,
and is peaked about a set of classical Lorentzian four-geometries. One may
then use the WKB approximation and the tangent vector to the configuration 
space -- for paths about which $\Psi$ is peaked -- to define the proper time
$\tau$ along the classical trajectories\cite{Halliwell}. Thus, once the mechanism of 
spontaneous breaking of the spacetime symmetries has taken place
in the early Universe, then the problems of time and time's arrow are solved
by means of the classical Hamilton-Jacobi equation and the classical
trajectories define a time and time's direction in the symmetric phase.
The Universe is then clearly divided into
a quantum gravity regime and a classical regime, making the WKB solution
to the origin of the time variable unambiguous without arbitrary boundary
conditions.

In the following, we shall investigate further the physical
consequences of the spontaneous breaking of the symmetries of spacetime
in the early Universe. We shall postulate that the velocity of light
goes through a first order phase transition at a critical time $t\sim t_c$,
associated with the spontaneous Higgs breaking of local Lorentz invariance.
This defines a ``superluminary'' phase of the early Universe, which solves
the horizon problem, the flatness problem and suppresses any relic
magnetic monopoles. In Section 4, the generation of quantum fluctuations in the
superluminary Universe is studied, and a scale invariant gaussian 
fluctuation density amplitude is derived from a model of the 
the critical phase transition.

\section{\bf  Spontaneous Breaking of Spacetime Symmetries}

Let us define the vierbein $e^a_\mu$ in terms of the metric: 
\begin{equation}
g_{\mu\nu}(x)=e^a_{\mu}(x)e^b_{\nu}(x)\eta_{ab}.
\end{equation}
The vierbeins $e^a_{\mu}$ satisfy the orthogonality relations:
\begin{equation}
e^a_{\mu}e_b^{\mu}=\delta^a_b,\quad e^{\mu}_ae_{\nu}^a=\delta^{\mu}_{\nu},
\end{equation}
which allow us to pass from the flat tangent space coordinates (the fibre
bundle tangent space) labeled by $a,b,c...
$ to the the world spacetime coordinates (manifold) labeled by $\mu,\nu,
\rho...$. 
The fundamental form (1) is invariant under Lorentz transformations:
\begin{equation}
e^{\prime\,a}_{\mu}(x)=L^a_b(x)e^b_{\mu}(x),
\end{equation}
where $L^a_b(x)$ are the homogeneous $SO(3,1)$ Lorentz transformation 
coefficients that can depend on position in spacetime, and which satisfy
\begin{equation}
L_{ac}(x)L^a_d(x)=\eta_{cd}.
\end{equation}

The $e^a_\mu$ obey the equation:
\begin{equation}
D_{\sigma}e^a_{\mu}=e_{\mu,\sigma}^a+(\Omega_{\sigma})^a_ce^c_{\mu},
\end{equation}
where $D_\sigma$ is the covariant derivative operator with respect to the 
(spin) gauge connection $\Omega_{\mu}$. 

In holonomic coordinates, the curvature tensor is given by
\begin{equation}
{R^{\lambda}}_{\sigma\mu\nu}=(R_{\mu\nu})^a_be^{\lambda}_ae^b_{\sigma}
\end{equation}
and the scalar curvature takes the form
\begin{equation}
R=e^{\mu a}e^{\nu b}(R_{\mu\nu})_{ab}.
\end{equation}

The action for Einstein gravity is written as 
\begin{equation}
S_G=-{c^4\over 16\pi G}\int d^4x e(R+2\Lambda),
\end{equation}
where $e\equiv \sqrt{-g}=\hbox{det}(e^a_{\mu}e_{a\nu})^{1/2}$ and $\Lambda$
is the cosmological constant.

We shall consider a specific
kind of symmetry breaking in the early Universe, in which the local Lorentz
vacuum symmetry is spontaneously broken by a Higgs mechanism.
We postulate the existence of four scalar fields, $\phi_a$, and assume that the 
vacuum expectation value (vev) of the scalar fields, $<\phi_a>_0$, will vanish 
for some temperature T less than a critical temperature $T_c$, when 
the local Lorentz symmetry is restored. Above $T_c$ the non-zero
vev will break the symmetry of the gound state of the Universe from $SO(3,1)$ 
down to $O(3)$. The domain formed by the direction of the vev of the Higgs field
will produce a time arrow pointing in the direction of increasing
entropy and the expansion of the Universe.
The four real scalar fields $\phi^a(x)$ are invariant
under Lorentz transformations\cite{Moffat}:
\begin{equation}
\phi^{\prime\,a}(x)=L^a_b(x)\phi^b(x).
\end{equation}
We can use the vierbein to convert $\phi^a$ into a 4-vector in coordinate
space: $\phi^{\mu}=e^{\mu}_a\phi^a$.
The covariant derivative operator acting on $\phi^a$ is defined by
\begin{equation}
D_{\mu}\phi^a=[\partial_{\mu}\delta^a_b+(\Omega_{\mu})^a_b]\phi^b.
\end{equation}

We shall now introduce a Higgs sector into the Lagrangian density such
that the gravitational vacuum symmetry, which we set equal to the Lagrangian 
symmetry at low temperatures, will break to a smaller symmetry at high
temperature. The pattern of vacuum phase transition that emerges contains
a symmetry anti-restoration[5-12]. This vacuum symmetry breaking leads
to the interesting possibility that exact zero temperature conservation laws
e.g. electric charge and baryon number are broken in the early
Universe. In our case, we shall find that the spontaneous breaking of the
Lorentz symmetry of the vacuum leads to a violation of the exact zero 
temperature conservation of energy in the early Universe.

Let us consider the Lorentz invariant Higgs potential:
\begin{equation}
V(\phi)=-{1\over 2}\mu^2\sum_{a=0}^3\phi_a\phi^a+\lambda\sum_{a=0}^3
(\phi_a\phi^a)^2,
\end{equation}
where we choose $\lambda > 0$, so that the potential is bounded from below.
Our Lagrangian density now takes the form:
\begin{equation}
{\cal L}={\cal L}_G +\sqrt{-g}\biggl[{1\over 2}D_{\mu}\phi_aD^{\mu}\phi^a
-V(\phi)\biggr].
\end{equation}
If $V$ has a minimum at $\phi_a=v_a$, then the spontaneously broken solution
is given by $v_a^2=\mu^2/4\lambda$ and an expansion of $V$ around the 
minimum yields the mass matrix:
\begin{equation}
(\mu^2)_{ab}={1\over 2}\biggl({\partial^2 V\over \partial \phi_a \partial 
\phi_b}\biggr)_{\phi_a=v_a}.
\end{equation}
We can choose $\phi_a$ to be of the form
\begin{equation}
\phi_a=\left(\matrix{0\cr 0\cr 0\cr v\cr}\right)
=\delta_{a0}(\mu^2/4\lambda)^{1/2}.
\end{equation}

All the other solutions of $\phi_a$ are related to this one by a Lorentz
transformation. Then, the homogeneous Lorentz group $SO(3,1)$ is broken 
down to the
spatial rotation group $O(3)$. The three rotation generators $J_i
(i=1,2,3)$ leave the vacuum invariant
\begin{equation}
J_iv_i=0,
\end{equation}
while the three Lorentz-boost generators $K_i$ break the vacuum symmetry
\begin{equation}
K_iv_i\not= 0.
\end{equation}
The $J_i$ and $K_i$ satisfy the usual commutation relations
\begin{equation}
[J_i,J_j]=i\epsilon_{ijk}J_k,\quad [J_i,K_j]=i\epsilon_{ijk}K_k,\quad
[K_i,K_j]=-i\epsilon_{ijk}K_k.
\end{equation}
The mass matrix $(\mu^2)_{ab}$ can be calculated from (13):
\begin{equation}
(\mu^2)_{ab}=(-{1\over 2}\mu^2+2\lambda v^2)\delta_{ab}+4\lambda v_av_b
=\mu^2\delta_{a0}\delta_{b0},
\end{equation}
where $v$ denotes the magnitude of $v_a$. There are three zero-mass Goldstone
bosons, the same as the number of massive vector bosons, and there are three
massless vector bosons corresponding to the unbroken $O(3)$ symmetry. In 
addition to these particles, one massive physical boson particle 
$h$ remains, after the spontaneous breaking of the vacuum. The vector boson 
mass term is given in the Lagrangian by
\begin{equation}
{\cal L}_{\Omega}={1\over 2}\sqrt{-g}(\Omega_{\mu})^{ab}v_b
(\Omega^{\mu})_a^cv_c
={1\over 2}\sqrt{-g}\sum_{i=1}^3((\Omega_{\mu})^{i0})^2(\mu^2/4\lambda).
\end{equation}

A phase transition is assumed to occur at the critical temperature $T_c$,
when $v_a\not= 0$ and the Lorentz symmetry is broken and the three gauge
fields $(\Omega_{\mu})^{i0}$ become massive vector bosons. Below
$T_c$ the Lorentz symmetry is restored, and we
regain the usual classical gravitational field with massless gauge fields
$\Omega_{\mu}$. The symmetry breaking will extend to the 
singularity or the possible singularity-free initial state of the big bang,
and since quantum effects associated with gravity do not become important
before $T\sim 10^{19}$ GeV, we expect that $T_c\leq 10^{19}$ GeV.

In most known cases of phase transitions of the first and second kind, the more 
symmetrical phase corresponds to higher temperatures and the less symmetrical
one to lower temperatures. A transition from an ordered
to a disordered state usually occurs with increasing temperature. Examples of
two known exceptions in Nature are the ``lower Curie point" of Rochelle salt,
below which the crystal is orthorhombic, but above which it is monoclinic.
Another example is the gapless superconductor.
A calculation of the effective potential for the Higgs breaking contribution in
(12) shows that extra minima in the potential $V(\phi)$ can occur for a 
noncompact group such as $SO(3,1)$\cite{Skagerstam}.

The entropy will rapidly increase during the phase transition, when the
symmetry is restored, and for a closed Universe
will reach a maximum at the final singularity, provided that no further
phase transition occurs which breaks the Lorentz symmetry of the vacuum. 
Thus, the symmetry breaking
mechanism explains in a natural way the low entropy at the initial
singularity and the large entropy at the final singularity. 
Since the
ordered phase is at a much lower entropy than the disordered phase and due
to the existence of a domain determined by the direction of the vev of the
fields, $<\phi_a>_0$, a natural explanation is given for the cosmological 
arrow of time and the origin of the second law of thermodynamics. 
Thus, the spontaneous symmetry breaking of the gravitational
vacuum corresponding to the breaking pattern, $SO(3,1)\rightarrow O(3)$, 
leads to a manifold with the structure $R\times O(3)$, in which time 
appears as an absolute external parameter. The vev of the fields,
$<\phi_a>_0$, points in a chosen direction of time to break the symmetry
creating an arrow of time. 

The total action for the theory is
\begin{equation}
S=S_G+S_M+S_{\phi},
\end{equation}
where $S_G$ is given by (8) and $S_M$ is the usual matter action for gravity.
Moreover,
\begin{equation}
S_{\phi}=\int d^4x\sqrt{-g}[{1\over 2}(D_{\mu}\phi_a
D^{\mu}\phi^a-V(\phi)].
\end{equation}

Performing a variation of $S$ leads to the field equations:
\begin{equation}
G^{\mu\nu}\equiv R^{\mu\nu}-{1\over 2}g^{\mu\nu}R={8\pi G\over c^4}(T^{\mu\nu}
+C^{\mu\nu})+\Lambda g^{\mu\nu},
\end{equation}
where $T^{\mu\nu}$ is the energy-momentum tensor for matter.
The energy-momentum tensor for the coordinate scalar $\phi_a$ fields 
is of the usual form:
\begin{equation}
C^{\mu\nu}=D^{\mu}\phi_a D^{\nu}\phi^a - {\cal L}_{\phi}g^{\mu\nu}.
\end{equation}

The Higgs mass and graviton
mass contributions proportional to $v^a=<\phi^a>_0$ are given by
(18) and (19).
Since we assume that the symmetry breaking pattern is $SO(3,1)\rightarrow O(3)$, 
there will be three massless gauge vector fields $(\Omega_{\mu})_{nm}
=-(\Omega_{\mu})_{mn}$, denoted by $U^n_{\mu}$, three massive vector
bosons $(\Omega_\mu)_{0i}=-(\Omega_\mu)_{i0}$, denoted by $V_{\mu}^i$ and one
massive boson $h$. 
Because $G^{\mu\nu}$ satisfies the Bianchi identities ${G^{\mu\nu}}_{;\nu}=0$,
we find in the broken symmetry phase:
\begin{equation}
{T^{\mu\nu}}_{;\nu}=K^\mu,
\end{equation}
where ; denotes the covariant derivative with respect to the Christoffel
symbols $\Gamma^{\lambda}_{\mu\nu}$, and $K^\mu$ contains the mass terms
proportional to $v_a=<\phi_a>_0$.
Thus, the conservation of energy-momentum is spontaneously violated and
matter can be created in this broken symmetry phase.
When the temperature passes below the critical temperature,
$T_c$, then $v_a=0$ and the
action is restored to its classical form with a symmetric degenerate 
vacuum and a massless spin gauge connection $(\Omega_{\mu})^a_b$, and 
we regain the standard 
energy-momentum conservation laws: ${T^{\mu\nu}}_{;\nu}=0$.

\section {\bf Field Equations in the Broken Symmetry Phase}

The ``Newtonian" spacetime manifold in the broken phase has the symmetry 
$R\times O(3)$. The three-dimensional space with $O(3)$ symmetry
is assumed to be homogeneous and isotropic and yields the usual
maximally symmetric three-dimensional space:
\begin{equation}
d\sigma^2=R^2(t)\biggl[{dr^2\over 1-kr^2}+r^2(d\theta^2
+\hbox{sin}^2\theta d\phi^2)\biggr],
\end{equation}
where $t$ is the external
time variable. This is the Robertson-Walker theorem for our ordered phase
of the vacuum and it has the correct subspace structure for the FRW 
Universe with the metric:
\begin{equation}
ds^2=dt^2c^2-R^2(t)\biggl[{dr^2\over 1-kr^2}+r^2(d\theta^2
+\hbox{sin}^2\theta d\phi^2)\biggr].
\end{equation}
The null geodesics of the metric (26) are the light paths of the subspace
and $d\lambda=ds/c$ measures the time at each test 
particle\cite{Rindler}, where $c$ is a constant with the dimensions of the speed of
light.

In the broken symmetry phase, the ``time" $t$ is the {\it absolute physical 
time} measured by standard clocks. In contrast to GR, while $<\phi>$ is 
non-zero, we no longer have re-parameterization invariance and time is no 
longer an arbitrary label.

The total action for the theory in the broken symmetry phase, $T>T_c$, is
\begin{equation}
S=S_G+S_M+S_h+S_V.
\end{equation}
Let us consider small oscillations about the true minimum and define a shifted
field:
\begin{equation}
\phi^{\prime}_a=\phi_a-v_a.
\end{equation}
After Faddeev-Popov ghost fixing, we can define a new set of 
extended Becchi-Rouet-Stora (BRS)\cite{Becchi} Lorentz gauge transformations 
under which $S$ is invariant, and a set of Ward-Takahashi identities 
can be found. 
Let us perform a Lorentz transformation on $\phi^a$, so that we obtain:
\begin{equation}
\phi^0=h,\quad \phi^1=\phi^2=\phi^3=0.
\end{equation}
In this special
coordinate frame, the remaining component $h$ is the physical Higgs
particle that survives after the Goldstone modes have been removed. This
corresponds to choosing the ``unitary gauge'' in the standard electroweak
model\cite{Halzen}.
In our specially chosen coordinate frame in which (29) holds and the metric
(26) is realized, we have
\begin{equation}
S_h=\int d^4x\sqrt{-g}[{1\over 2}\partial_{\mu}h
\partial^{\mu}h-V(h)],
\end{equation}
where
\begin{equation}
V(h)=4\lambda v^2h^2+4\lambda vh^3+\lambda h^4-{1\over 2}V_\mu^2h^2-vV_\mu^2h,
\end{equation}
and we have for convenience suppressed the index $i$ on $V_\mu^i$. Moreover,
\begin{equation}
S_V={1\over 2}m^2\int d^4x\sqrt{-g}g^{\mu\nu}V_{\mu}V_{\nu},
\end{equation}
where the mass $m\propto <h>$. 

The field equations are of the form:
\begin{equation}
G^{\mu\nu}\equiv R^{\mu\nu}-{1\over 2}g^{\mu\nu}R={8\pi G\over c^4}
(T^{\mu\nu}+K^{\mu\nu}+H^{\mu\nu})+\Lambda g^{\mu\nu},
\end{equation}
where $K^{\mu\nu}$ is given by
\begin{equation}
K^{\mu\nu}=m^2(V^{\mu}V^{\nu}-{1\over 2}g^{\mu\nu}V^{\beta}V_{\beta}).
\end{equation}
Moreover, the $h$ field energy-momentum tensor is of the usual form:
\begin{equation}
H^{\mu\nu}=\partial^{\mu}h \partial^{\nu}h - {\cal L}_h g^{\mu\nu}.
\end{equation}

The spin connections $(\Omega_\sigma)^{ab}$ are determined by the equations:
\begin{equation}
e^a_{\mu,\sigma}+(\Omega_\sigma)^a_ce^c_\mu-\Gamma^\rho_{\sigma\mu}e^a_\rho=0,
\end{equation}
which leads to the solution:
\begin{equation}
(\Omega_\sigma)^{ab}=-e^{b\mu}(e^a_{\mu,\sigma} -     
e^a_\rho\Gamma^\rho_{\sigma\mu}).
\end{equation}
For an FRW manifold, we have
\begin{equation}
e^0_\mu=\delta^0_\mu,\quad e^1_\mu={R\over (1-kr^2)^{1/2}}\delta^1_\mu,
\end{equation}
and
\begin{equation}
\quad e^2_\mu=Rr\delta^2_\mu,\quad e^3_\mu=Rr\hbox{sin}\theta\delta^3_\mu.
\end{equation}

Moreover, we have
\begin{equation}
\Gamma^k_{00}=\Gamma^0_{k0}=\Gamma^0_{00}=0,\quad \Gamma^i_{kl}=
{1\over 2}\gamma^{ij}(\gamma_{jl,k}+\gamma_{jk,l}-\gamma_{kl,j}),
\end{equation}
and
\begin{equation}
\Gamma^0_{kl}={1\over c}R\dot R\gamma_{kl},\quad
\Gamma^k_{0l}={\dot R\over cR}\delta^k_l,
\end{equation}
where $\gamma_{ik}$ is the three-metric associated with (25) and
$\dot R=dR/dt$. The massive spin connection components  
$V_\sigma=(\Omega_\sigma)^{0i}$ are determined in our Newtonian FRW manifold by
\begin{equation}
(\Omega_0)^{i0}=(\Omega_0)^{0i}=0,\quad (\Omega_k)^{i0}=e^i_k{\dot R\over cR}.
\end{equation}

Since $G^{\mu\nu}$ satisfies the Bianchi identities:
\begin{equation}
{G^{\mu\nu}}_{;\nu}=0,
\end{equation}
we find from (33) that
\begin{equation}
{T^{\mu\nu}}_{;\nu}=-(K^{\mu\nu}+H^{\mu\nu})_{;\nu},
\end{equation}
where we have used ${g^{\mu\nu}}_{;\nu}=0$.
In the unbroken phase of spacetime, we regain the standard energy-momentum
conservation laws ($K^{\mu\nu}=0$):
\begin{equation}
{T^{\mu\nu}}_{;\nu}={C^{\mu\nu}}_{;\nu}=0.
\end{equation}

\section{\bf Superluminary Universe}

Let us consider a locally flat patch of space in the broken symmetry phase
for $t<t_c$, with the line element:
\begin{equation}
ds^2=c_0^2dt^2-(dx^1)^2-(dx^2)^2-(dx^3)^2,
\end{equation}
where $c_0$ is the value of the velocity of light in the broken symmetry 
phase when $<\phi>\not= 0$.
From the geodesic solution for light travel, $ds^2=0$, we get
\begin{equation}
dt^2={1\over c_0^2}[(dx^1)^2+(dx^2)^2+(dx^3)^2].
\end{equation}
In the limit $c_0\rightarrow \infty$, $dt\rightarrow 0$ and the Minkowski light 
cone is squashed.
This solves the horizon problem, since all the points in an expanding bubble
near the beginning of the Universe will be in communication with one another.

The horizon scale is determined by
\begin{equation}
d_H(t)=c_0R(t)\int_0^t{dt^{\prime}\over R(t^{\prime})}.
\end{equation}
For $t > t_c$, this will have the usual value: $d_H(t) = 2ct$, 
since $R(t)\propto t^{1/2}$ for a radiation dominated Universe. 
Let us assume that for $t \leq t_c$, the speed of light is very
large. During a first order phase transition, the velocity of light is 
assumed to undergo a discontinuous change from, for example, the value:
\begin{equation}
c_0\sim 5\times 10^{28}c
\end{equation}
for $t\leq t_c$ to $c_0=c$ (c is the present value of the velocity of 
light) for $t > t_c$, as shown in Fig. 1a. Then, we get for $t\leq t_c$:
\begin{equation}
d_H(t)\approx c_0g(t),
\end{equation}
where $g(t)$ is the functional time dependence arising from $R(t)$ in
(48). Thus, for a fraction of time near the beginning of the Universe, all 
points of the expanding space will have been in communication with one another,
solving the horizon problem. 

Suppose the Higgs field that breaks the spacetime symmetries is characterized
by a correlation length $\xi$. Then, the monopole density is approximately
given by
\begin{equation}
n_M\approx \xi^{-3}.
\end{equation}
In the superluminary model, the bound on the length $\xi$ is given by
\begin{equation}
\xi < d_H(t) \approx c_0g(t),
\end{equation}
so that the bound on the number density of monopoles is exponentially
weakened. This solves the relic particle (monopole) problem.

For the standard FRW models the quantity $\Omega_0$, along with $H_0$ 
($H=\dot R/R$),
determines the cosmological model, for $\Omega_0$ and $H_0$ 
determine the radius of curvature: 
\[
R^2=\break (c^2/H_0^2)/\vert\Omega_0-1\vert,
\]
and the density: $\rho_0=(3H_0^2/8\pi G)\Omega_0$. The present observational data restrict
$\Omega_0$ to lie in the interval [0.01, few], which implies that
$R\sim c/H_0$ and $\rho_0\sim \rho_{\hbox{crit}}$. Now consider in the unbroken
phase the expression\cite{Turner}:
\begin{equation}
\Omega(t)=1/[1-x(t)],
\end{equation}
where
\begin{equation}
x(t)=(kc^2/R^2)/(8\pi G\rho/3).
\end{equation}
Then, we have
\begin{equation}
\vert \Omega(10^{-43}\hbox{sec})-1\vert \leq O(10^{-60}),
\end{equation}
and
\begin{equation}
\vert \Omega(1\hbox{sec})-1\vert\sim O(10^{-16}).
\end{equation}
Thus, the standard FRW cosmology is characterized by the very special initial 
data:
\begin{equation}
\vert \Omega-1\vert\, \leq O(10^{-60}).
\end{equation}
This is the source of the flatness problem. 

Let us assume, as an example, that in the spontaneously broken phase for 
$t_c\approx 10^{-35}$ s, the velocity of light has the large value (49).
The initial data at the Planck epoch are now:
\begin{equation}
\vert \Omega-1\vert \leq O(10^{-3}),
\end{equation}
and
\begin{equation}
\vert (kc_0^2/R^2)\vert/(8\pi G\rho/3) \sim O(10^{-3}),
\end{equation}
which implies much less fine tuning than in the standard FRW model
of the Universe. Clearly, the value of $c_0$ is bounded by 
$c_0 < 10^{30}c$ to prevent an early collapse of the Universe. 

We observe that in contrast to 
the inflationary model (which predicts that $\Omega=1$), the prediction for the 
value of $\Omega$ in the superluminary model depends on the detailed dynamics
of the theory. Indeed, if we were to assume the equation of state: 
$\rho=$ const. in the broken phase for 
which $R(t)$ has the inflationary behavior: $R(t)\propto 
\hbox{exp}(Ht)$,
then we would regain the standard inflationary prediction $\Omega=1$. 

\section{\bf Quantum Fluctuations}

We can allow the possibility in our model for generating the seed
perturbations that can grow to form the large-scale structures.
In the FRW model with $R(t)\propto t^{1/2}$, the physical 
wavelengths, which grow as $\lambda\propto R(t)\propto t^{1/2}$, will be far
larger than the horizon (which grows as $cH(t)^{-1}\sim ct$) in the early
phases. The comoving scales $\lambda$ cross the horizon only once. They
start larger than the horizon, i.e. $\lambda \gg cH^{-1}$, and then cross
inside the horizon at a later epoch. During the superluminary phase for 
$t < t_c$, the fluctuation wavelengths grow as: $\lambda\propto R(t)$.
However, the horizon grows rapidly, $d_H\approx c_0g(t)$, where $c_0$ is
given by (49), and will become equal to
the physical wavelength at some time $t=t_{\hbox{exit}}$, after which
it becomes larger than $\lambda(k)$ for a mode labeled by a wave vector 
${\vec k}$. 

After the symmetry is restored at
$t > t_c$, the proper length $R(t)$ grows as $t^{1/2}$, 
whereas the horizon will increase as $cH(t)^{-1}\sim ct$. Therefore, the
wavelength will be completely within the Hubble radius for an interval of time
$\Delta t$.  Thus, in the superluminary model the fluctuations are 
in microcausal connection very early in the Universe ($t\sim 10^{-35}$ s)
and have time to grow
into physical modes sufficiently large to form Galaxy structures. These
fluctuations will have a gaussian form, provided any self-couplings of the 
matter fields are small.

Quantum fluctuations in matter fields taking place in the epoch, $t <
t_c$, are candidates for seed perturbations. For example,
the fluctuations associated with the Higgs field, $h$, which are
responsible for breaking the spacetime symmetries could be such a candidate,
satisfying the equation:
\begin{equation}
\ddot h+3H\dot h+ \Gamma_h\dot h+V^{\prime}(h)=0,
\end{equation}
where $V(h)$ is given by (31),
$\Gamma_\phi$ is the decay width associated with the decay of the 
$h$ particle and $V^{\prime}(h)=dV(h)/dh$. 

In inflationary models, the Hubble parameter is constant during inflation, 
so all interesting scales begin sub-horizon, cross outside the Hubble radius
during inflation, and at a later epoch cross back inside the horizon (at
the usual time for standard FRW models). As a result, the spectrum of
fluctuations predicted is close to a scale invariant spectrum--the 
Harrison-Zel'dovich spectrum\cite{Harrison}. 
If we assume in the superluminary model that $c$ undergoes a discontinuous 
change, during a first order phase transition at $t\sim
t_c$, to the value $c_0$ given by (49),
then the horizon, $d_H(t)$, determined by (48) will also
have a discontinuity in its first derivative with respect to $t$, and
$d_H(t)$ for $t < t_c$ can be matched to $d_H(t)$ for $t 
\geq t_c$ in such a way that
$\lambda$ crosses $d_H(t)$ twice, as in the inflationary model.
The fluctuations
are ``frozen in'' and leave an imprint on the metric tensor. 

With the assumption that $h$ is spatially homogeneous, $H^{\mu\nu}$ takes the
form of a perfect fluid, with the energy density and pressure given by
\begin{equation}
\rho_h={1\over 2}\dot h^2+V(h),\quad c^2p_h={1\over 2}\dot h^2-V(h).
\end{equation}
The fluctuations of the field $h$ are defined by the Fourier transform:
\begin{equation}
\delta h_k=\int d^3x \hbox{exp}(i\vec k\cdot \vec x)\delta h(x),
\end{equation}
and we have
\begin{equation}
(\Delta h)^2_k=V^{-1}k^3\vert \delta h_k\vert^2/2\pi^2
=\biggl({H\over 2\pi}\biggr)^2.
\end{equation}
Fluctuations in $h$ give rise to perturbations in the energy density:
\begin{equation}
\delta \rho_h=\delta h\biggl({\partial V\over \partial h}\biggr).
\end{equation}

At horizon crossings, $\lambda_{\hbox{phys}}\sim cH^{-1}$, the gauge invariant
quantity $\zeta$ takes the simple form $\zeta=\delta\rho/(\rho+c^2p)\,$\cite{Bardeen}.
In the radiation dominated era and in the matter dominated era, $\zeta$ at
horizon crossing is, up to a factor of order unity, equal to 
$\delta \rho/\rho$. Equating the values of $\zeta$ at the two horizon 
crossings, we find
\begin{equation}
\biggl({\delta \rho\over \rho}\biggr)_{\hbox{Hor}}
\sim {\delta h V^{\prime}\over \dot h^2}\sim
{HV^{\prime}\over 2\pi \dot h^2},
\end{equation}
where we have used the fact that $\delta h\sim H/2\pi$. 
We must now model $V^{\prime}$ and $\dot h$ at the phase transition, in order
to estimate the density fluctuation, $\delta \rho/\rho$. Clearly, 
in contrast to the inflationary models, $H$ is
rapidly varying at the phase transition. We have
\begin{equation}
\dot h\sim {H\over 2\pi \delta t}.
\end{equation}
A natural time scale for the duration of the phase transition is given by
\begin{equation}
\delta t\sim \biggl({H\over 2\pi h^3}\biggr)^{1/2}.
\end{equation}
Thus, if we choose $1/H\sim 10^{-34}$ s and $h\sim M_P\sim 10^{43} s^{-1}$, 
then the duration of the phase transition is $\delta t\sim 10^{-48}$ s. 

By assuming that $V(h)$, in (31), is dominated by $V(h)
\sim {\lambda\over 4} h^4$, we obtain from (64) the scale invariant 
prediction for the amplitude:
\begin{equation}
\biggl({\delta\rho\over \rho}\biggr)_{\hbox{Hor}}\sim \lambda.
\end{equation}
We can fit the data measured by
the Cosmic Background Explorer (COBE), which is consistent with a gaussian,
scale invariant spectrum\cite{Smoot}, by choosing the coupling constant 
$\lambda\sim 10^{-5}$, and using 
$\Delta T/T\sim {1\over 3}\delta\rho/\rho$. The measurements are quoted in terms
of a spectral index $n$, with $n=1.1\pm 0.5$. These measurements are
also consistent with the predictions of inflationary models, and with other
mechanisms of inhomogeneity generation, such as cosmic strings.

In standard inflationary models, the COBE data constrains the coupling constant
of the inflaton models to be, $\lambda\sim 10^{-12}-10^{-14}$, which seems
unnaturally small for the physical models considered in inflationary
scenarios.

Another interesting feature of the superluminary model is that the
Planck length, $L_P=(\hbar G/c_0^3)^{1/2}\rightarrow 0$ as $c_0\rightarrow
\infty$ or, alternatively, the Planck mass, $M_P=\break (\hbar c_0/G)^{1/2}
\rightarrow\infty$ in this limit. This could have important implications for
quantum fluctuations and for quantum gravity in the symmetry broken phase.

The results obtained above suggest that the
superluminary model could be an attractive alternative to inflation as a
solution to the initial value problem in cosmology.
\vskip 0.3 true in

{\bf Acknowledgments}
\vskip 0.2 true in
I thank M. Clayton, N. Cornish, P. Savaria, G. Starkman and 
D. Tatarski for helpful and stimulating discussions. This work was supported by the Natural
Sciences and Engineering Research Council of Canada.
\vskip 0.5 true in

\end{document}